\documentclass[usenatbib]{mnras}

\usepackage[T1]{fontenc}
\usepackage{ae,aecompl}

\usepackage{amsmath}
\usepackage{graphicx}
\usepackage{amsmath}	
\usepackage{amssymb}	
\PassOptionsToPackage{authoryear}{natbib}
\usepackage{natbib}
\bibpunct{(}{)}{;}{a}{}{,}
\PassOptionsToPackage{american}{babel}
\usepackage{babel}

\newcommand\arcdeg{\ensuremath{\degr}}
\newcommand\arcm{\ensuremath{^\prime}}
\newcommand{\pass}{{\sffamily{Pass 8}}}
\PassOptionsToPackage{hyphens}{url} 

\title[A cut-off in the TeV gamma-ray spectrum of the SNR Cassiopeia A]{A cut-off in the TeV gamma-ray spectrum of the SNR Cassiopeia A}

\author[MAGIC Collaboration]{
  \parbox{\linewidth}{
MAGIC Collaboration:
    M.~L.~Ahnen$^{1}$,
S.~Ansoldi$^{2,21}$,
L.~A.~Antonelli$^{3}$,
C.~Arcaro$^{4}$,
A.~Babi\'c$^{5}$,
B.~Banerjee$^{6}$,
P.~Bangale$^{7}$,
U.~Barres de Almeida$^{7}$,
J.~A.~Barrio$^{8}$,
J.~Becerra Gonz\'alez$^{9}$
% ...
\\
(Affiliations can be found after the references)
  }
}

\begin{document}

\maketitle

\clearpage

\date{Accepted XXX. Received YYY; in original form ZZZ}

\pubyear{2017}

\pagerange{\pageref{firstpage}--\pageref{lastpage}}
\label{firstpage}

\begin{abstract}
It is widely believed that the bulk of the Galactic cosmic rays are accelerated in supernova remnants (SNRs). However, no observational evidence of the presence of particles of PeV energies in SNRs has yet been found. The young historical SNR Cassiopeia A (Cas A) appears as one of the best candidates to study acceleration processes. Between December 2014 and October 2016 we observed Cas A with the MAGIC telescopes, accumulating 158 hours of good-quality data. We derived the spectrum of the source from 100 GeV to 10 TeV. We also analysed $\sim$8 years of $Fermi$-LAT to obtain the spectral shape between 60 MeV and 500~GeV. The spectra measured by the LAT and MAGIC telescopes are compatible within the errors and show a clear turn off (4.6~$\sigma$) at the highest energies, which can be described with an exponential cut-off at $E_c = 3.5\left(^{+1.6}_{-1.0}\right)_{\textit{stat}} \left(^{+0.8}_{-0.9}\right)_{\textit{sys}}$~TeV. The gamma-ray emission from 60 MeV to 10~TeV can be attributed to a population of high-energy protons with spectral index $\sim$2.2 and energy cut-off at $\sim$10~TeV. This result indicates that Cas A is not contributing to the high energy ($\sim$PeV) cosmic-ray sea in a significant manner at the present moment. A one-zone leptonic model fails to reproduce by itself the multi-wavelength spectral energy distribution. Besides, if a non-negligible fraction of the flux seen by MAGIC is produced by leptons, the radiation should be emitted in a region with a low magnetic field (B$\lessapprox$100$\mu$G) like in the reverse shock.
\end{abstract}
   



%

\section{Introduction}\label{sect:intro}

Supernova remnants (SNRs) are widely believed to be able to accelerate cosmic rays (CRs) to PeV energies, and of being the main contributors to the galactic CR sea \citep{Berezhko_2003,bell_2013_2,Drury_2014}. 
Two arguments support this belief. On one hand SNRs can explain the observed energy density of CRs if one assumes that around 10\% of the kinetic energy of the supernova (SN) explosion goes into CR acceleration and a supernova explosion rate of $\sim$3 per century \citep{ginzburg_1964,gaisser_1991}. On the other hand Diffusive Shock 
Acceleration (DSA,~\citealt{bell_2013}) offers a plausible acceleration mechanism and explains the CR spectral shape. 
Observations at high and very high energies have further strengthened
this paradigm: several SNRs have been observed to emit at TeV energies, a signature that particles are being accelerated to relativistic energies. The spectral shape observed in some SNRs at sub-GeV energies points to neutral pion decay as the origin of the high-energy emission~\citep{ackermann_2013, aharonian_2013}, however, the origin of the emission at the highest energies, in the TeV regime, is still uncertain.

If SNRs are the sources of all galactic CRs they must be able to accelerate particles all the way up to the \textit{knee} of the CR spectrum, a feature observed at around 3 PeV. In fact this represented an 
important theoretical challenge for decades, because standard DSA was unable to explain acceleration beyond 100 TeV. It has been later realised~\citep{bell_2004} that the magnetic field upstream of the shock of young SNR can be amplified due to instabilities produced by CRs themselves. The missing part to solve the paradox is the observational evidence: as of today no SNR has been found where hadronic CR acceleration up to PeV energies can be firmly established.

Cassiopeia A (Cas~A) is one of the few good candidates for these studies. The precise knowledge of the age of this core-collapse SNR (330 yrs), the remnant of a historical supernova in AD1680, allows the determination of many otherwise free parameters when studying its morphology and spectral shape. It is located at a distance of 3.4$^{+0.3}_{-0.1}$~kpc and has an angular diameter of 5 \arcm{}~\citep{reed_1995}.
It is the brightest radio source outside our solar system and it is in fact bright 
all over the electromagnetic spectrum, offering an excellent opportunity 
to study particle acceleration. 

Cas A  has been extensively observed in radio wavelengths \citep{Lastochkin_1963,Medd_1965,Allen_1967,Parker_1968,Braude_1969, Hales_1995}. Most of the emission comes from a bright radio ring of $\sim$1.7~pc radius and a faint outer plateau of $\sim$2.5~pc radius \citep{Zirakashvili_2014}, although a distinct emission coming from several compact and bright knots has also been identified \citep{Anderson_1991}. The spectral index of the radio flux can vary from $\sim$0.6 to $\sim$0.9 over the remnant. Several emission regions were also identified in the X-ray band \citep{Gotthelf_2001,suzaku_2009,Grefenstette_2015, integral_2016}. In the gamma-ray domain, $Fermi$-LAT detected the source at GeV energies~\citep{abdo_2010} and later derived a spectrum that displays a
low energy spectral break at 1.72$\pm$1.35 GeV~\citep{yuan_2013}. In the TeV energy range, Cas A was first detected by HEGRA~\citep{aharonian_2001} and later confirmed by MAGIC~\citep{albert_2007}. VERITAS has recently reported a spectrum extending well above 1~TeV~\citep{holder_2016}. The spectrum seems to steepen from the $Fermi$-LAT energy range to the TeV bands. The photon indices measured 
by HEGRA, MAGIC, and VERITAS are seemingly larger than the $Fermi$-LAT index of 2.17$\pm$0.09, but the statistical and systematic errors are too large for a firm conclusion.

Multi-wavelength modelling of Cas A observations has not yet resulted in a clear discrimination between hadronic and/or leptonic origin of the observed radiation in the GeV to TeV energy range (i.e. \citealt{Berezhko_2003,Vink_2003,yuan_2013,Saha_2014,Zirakashvili_2014}). However the break in the $Fermi$-LAT spectrum at $\sim$1~GeV combined with the observations at TeV energies suggest that the observed gamma-ray flux has either a pure hadronic origin or that several emission mechanisms (proton-proton interaction, inverse Compton and/or Bremsstrahlung) are involved. Indeed, several plausible acceleration regions have been identified in Cas A. Chandra X-ray images \citep{Gotthelf_2001} and high-resolution VLA radio synchrotron maps \citep{Anderson_1995} show a thin outer edge to the SNR that has been interpreted to represent the forward shock where the blast wave encounters the circumstellar medium \citep{DeLaney_2003}. The cold SNR ejecta expands supersonically outward from the explosion center producing a strong shock where the magnetic field can be amplified and hence accelerate CRs to PeV energies \citep{bell_2004,bell_2013}. This scenario was reinforced by the observations of year-scale variability in the synchrotron X-ray filaments of Cas A \citep{Uchiyama_2008}, which require a magnetic field amplification at the shock of the order of mG. High-resolution observations \citep{Gotthelf_2001,Morse_2004,Patnaude_2007,Helder_2008} also show a reverse shock formed well behind the forward shock that decelerates the impinging ejecta. The parameters that characterise the reverse shock can be significantly different from the ones describing the forward shock, enhancing different dominant radiation mechanisms on each zone. For instance, inverse Compton (IC) contribution, up-scattering the large FIR photon field of Cas A itself (with energy density of $\sim$2 eV/cm$^3$ and temperature of 97 K, \citealt{Mezger_1986}), is more significant in a region of lower magnetic field, as otherwise it would be suppressed due to fast cooling of electrons. Hard X-ray observations \citep{Grefenstette_2015,Siegert_2015}, if of synchrotron origin, prove the presence of relativistic electrons with Lorentz factor $\gamma_{e}\geq 100$, which can also produce gamma rays through relativistic bremsstrahlung.

We use the MAGIC telescopes to improve the accuracy of the spectral measurement at multi-TeV energies. We also derived the spectrum obtained with the LAT, selecting events with the best energy reconstruction, to extend the spectrum to lower energies and also have sufficient overlap at very high energies (VHE). The full spectrum obtained from $\sim$60~MeV to $\sim$10~TeV is investigated here to determine the underlying mechanisms powering the young remnant, constraining the maximum energy of the accelerated particles and their nature.

\section{Observations and data analysis}
\label{sect:observations}

\subsection{The MAGIC Telescopes}

MAGIC is a system of two 17~m diameter Imaging Atmospheric Cherenkov Telescopes (IACTs), located at an altitude of 2200~m a.s.l. at the Roque de los Muchachos Observatory on the Canary Island of La Palma, Spain (28$^\circ$N, 18$^\circ$W). The telescopes are equipped with photomultiplier tubes (PMTs) that can detect the flashes of Cherenkov light produced by extensive air showers initiated in the upper atmosphere by gamma-ray photons with energies $\gtrsim$50 GeV. In the absence of moonlight and for zenith angles less than $30^\circ$ MAGIC reaches an energy threshold of $\sim$50~GeV at trigger level, and a sensitivity above 220\,GeV of $0.67 \pm 0.04 \% $ of the Crab Nebula flux (C.U., \citealt{Magic_performanceII}). 

Observations were performed between December 2014 and October 2016, for a total observation time of 158 hours after data quality cuts. They were carried in the so-called wobble mode \citep{Fomin}, with a standard wobble offset of $0.4^\circ$. The data correspond to zenith angles between 28 and 50 degrees and most of them ($\sim$73\%) were taken during moonlight time (see Table \ref{tabTime}), under background-light levels that could be up to 12 times brighter than during dark nights. A significant part of the data ($\sim$24\%) were obtained under Reduced High Voltage (HV) settings: the gain of the PMTs is lowered by a factor $\sim$1.7 to decrease the damage inflicted by background light on the photodetectors during strong moonlight time. The main effect of moonlight in the performance of the telescopes is an increase in the energy threshold, which for zenith angles between 30 and 45 degrees goes from $\sim$100~GeV during dark conditions to $\sim$300~GeV in the brightest scenario considered. As achieving a low energy threshold was not critical for this project, Moon observations provided a unique way to accumulate observation time. A detailed study of the performance of the MAGIC telescopes under moonlight is reported in \citep{MAGIC_moon}.

The data have been analysed using the standard tools used for the analysis of the MAGIC telescope data, MARS~\citep{zanin2013mars} following the optimised moonlight analysis described in \citep{MAGIC_moon}. For the spectrum reconstruction a point-like source was assumed and typical selection cuts with 90\% and 75\% $\gamma$-ray efficiency for the $\gamma$-ray/hadron separation and sky signal region radius, respectively \citep{Magic_performanceII}. Three OFF regions were considered for the background estimation.

 \begin{table}
\centering
 \caption{Effective observation time of the different hardware and sky brightness conditions under wich Cas A samples were taken.}\label{tabTime}
\begin{tabular}{ c  c }
\hline
Observation conditions & Time [h] \\
\hline
   \hline

     Dark and Nominal HV & 42.2 \\
     Moon and Nominal HV & 77.7 \\     
     Moon and Reduced HV & 38.1 \\     
   \hline
   All configurations & 158.0 \\
 \end{tabular}
 \end{table}

\subsection{$Fermi$-LAT}

The GeV emission of Cas A was revisited using 3.7~yr of LAT observations \citep{yuan_2013}. The spectrum derived is well-represented by a broken power-law with a break of 6.9$\sigma$ significance at $\sim$1.7~GeV.
To compare with the observations performed with the MAGIC telescopes, and also to update and improve the spectrum, we analysed 8.3~yr of LAT data (up to December 6, 2016) on a 15\arcdeg{}$\times$15\arcdeg{} region around the position of Cas A\footnote{The analysis on a 30\arcdeg{}$\times$30\arcdeg{} region yields compatible results.}.
We selected events with energy between 60\,MeV and 500\,GeV and applied the usual filters and corrections recommended by the $Fermi$-LAT collaboration (removing intervals when the rocking angle of the LAT was greater than 52\arcdeg{} or when parts of the region-of-interest, ROI, were observed at zenith angles larger than 90\arcdeg{}, as well as enabling the energy dispersion). In order to derive the energy spectrum we applied a maximum likelihood estimation analysis in 12 independent energy bins from 60\,MeV to 500\,GeV, modelling the Galactic and isotropic diffuse emission using the templates provided by the $Fermi$-LAT collaboration\footnote{\emph{gll\_iem\_v06.fits} and \emph{iso\_P8R2\_ULTRACLEANVETO\_V6\_v06.txt},
{http://fermi.gsfc.nasa.gov/ssc/data/analysis/documentation/Cicerone}}. During the broad-band fit, all sources in the third $Fermi$ catalog (3FGL) within the ROI were included. A source located $\sim$3.7\arcdeg{} away from Cas A at (l,b)=(113.6\arcdeg{},1.1\arcdeg{}) was added during the fitting process to account for a significant residual excess (with TS$=45.08$). The spectral parameters of the
background sources were fixed to those previously found, except for the sources within 5\arcdeg{} of the candidate location and the normalisation of the two diffuse background components. Following the results obtained by \cite{yuan_2013} we used a smoothly broken power-law function to fit the broadband spectrum of Cas~A ($dN/dE = N_{\rm o}(\frac{E}{E_o})^{\Gamma_1}(1+(\frac{E}{E_b})^{(\Gamma_1-\Gamma_2)/\beta})^{-\beta}$)  with the parameter $\beta$ fixed to 1 and the energy break to $E_
{\rm b}$=1.7 GeV. E$_{\rm o}$ is the normalisation energy, fixed to 1~GeV. The data set was reduced and analysed using \emph{Fermipy}\footnote{http://fermipy.readthedocs.org/en/latest/}, a set of python tools which automatise the \pass{} analysis. 
We analysed the four EDISP event types separately and combined them later by means of a joint likelihood fit. 
The SED was obtained by fitting the source normalisation factor in each energy bin independently using a power-law spectrum with a fixed spectral index of 2. For each spectral point we required at least a TS of 4, otherwise upper limits at the 95\% confidence level were computed.

\section{Results}

\begin{figure}
\includegraphics[width=\columnwidth]{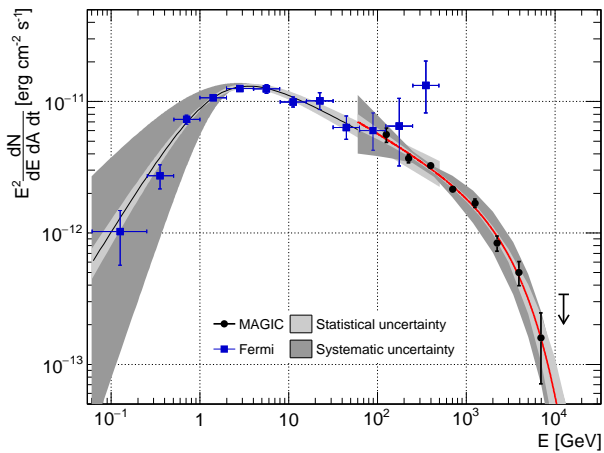}
\caption{Spectral energy distribution measured by the MAGIC telescopes (black dots) and $Fermi$ (blue squares). The red solid line shows the result of fitting the MAGIC spectrum with Eq. \ref{Eq:fit_EPWL}. The black solid line is the broken power-law fit applied to the $Fermi$ spectrum. 
}\label{fig:SED}
\end{figure}

Figure \ref{fig:SED} shows the reconstructed SED obtained with the MAGIC telescopes (black solid points). Red solid line is the curve obtained that best fits the MAGIC data assuming a power-law with an exponential cut-off (EPWL):

\begin{equation}\label{Eq:fit_EPWL}
\dfrac{dN}{dE} = N_0 \left( \dfrac{E}{E_0}\right)^{-\Gamma} \text{exp}\left(- \dfrac{E}{E_c} \right)
\end{equation}
with a normalisation constant ${N_0=(1.1 \pm 0.1_{\textit{stat}} \pm 0.2_{\textit{sys}})} \times 10^{-11}$~$\text{TeV}^{-1} \text{cm}^{-2} \text{s}^{-1}$ at a normalisation energy ${E_0 = 433~\text{GeV}}$, a spectral index ${\Gamma = 2.4 \pm 0.1_{\textit{stat}} \pm 0.2_{\textit{sys}}}$ and a cut-off energy $E_c = 3.5\left(^{+1.6}_{-1.0}\right)_{\textit{stat}} \left(^{+0.8}_{-0.9}\right)_{\textit{sys}}$~TeV. The spectral parameters of the tested models $\theta=\left\lbrace N_0,\Gamma,E_c \right\rbrace$ are obtained via a maximum likelihood approach. The data inputs are the numbers of recorded events (after background suppression cuts) in each bin of estimated energy $E^{i}_{\text{est}}$, both around the source direction ($N^{\text{ON}}_{i}$) and in the three OFF regions ($N^{\text{OFF}}_{i}$). An additional set of nuisance parameters $\mu_{i}$ for modelling the background are also optimized in the likelihood calculation. In each step of the maximisation procedure the expected number of gammas in a given bin of estimated energy ($E_{\text{est}}$) is calculated by folding the gamma spectrum with the MAGIC telescopes response (energy-dependent effective area and energy migration matrix). The background nuisance parameters and the statistical uncertainties in the telescopes response are treated as explained in \citep{Rolke_2005}.

The probability of the EPWL fit is 0.42. We tested the model against the null hypothesis of no cut-off, which is described with a pure power-law (PWL). The probability of the PWL fit is $6 \times 10^{-4}$. A likelihood ratio test between the two tested models favours the one that includes a cut-off at $\sim3.5$~TeV with 4.6$\sigma$ significance.

\begin{figure}
\includegraphics[width=\columnwidth]{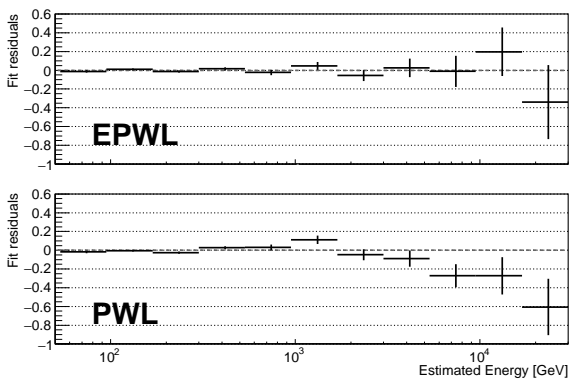}
\caption{Relative fit residuals for the two tested models fitting the MAGIC spectrum: power-law with exponential cut-off (EPWL, upper panel) and power law (PWL, lower panel). The error bars are calculated such that they correspond to the total contribution of each estimated energy bin to the final likelihood of the fit.}\label{fig:Res}
\end{figure}

Figure \ref{fig:Res} compares the fit residuals for the two tested models: PWL and EPWL. The residuals are here defined as $N^{\text{obs}}_{\text{ON}}/N^{\text{exp}}_{\text{ON}}-1$, where $N^{\text{obs}}_{\text{ON}}$ is the number of observed events (including background) in the ON region and $N^{\text{exp}}_{\text{ON}}$ is the number of events predicted by the fit in the same region. All the bins in estimated energy which contain events are used in the fits, but only those with a $2\sigma$ significance gamma-ray excess are shown as SED points in upper panel of Fig. \ref{fig:SED}.

The systematic uncertainty due to an eventual mismatch on the absolute energy scale between MAGIC data and MC simulations was constrained to be below 15\% in \cite{Magic_performanceII}. By conservatively modifying the absolute calibration of the telescopes by $\pm 15\%$, and re-doing the whole analysis, we can evaluate the effect of this systematic uncertainty in the estimated source spectrum. This does not produce a simple shift of the spectrum along the energy axis, but changes also its hardness. Even in the unlikely scenario in which, through the 158 h of observations, the {\it average} Cherenkov light yield was overestimated by 15\% relative to the MC, by applying the corresponding correction the resulting spectrum is still better fit by an EPWL at the level of 3.1$\sigma$. In the also unlikely scenario in which the light yield was underestimated, the EPWL is preferred over the PWL at the 6.5$\sigma$ level.
The systematic uncertainties in the flux normalization and spectral index were retrieved from the publication reporting the performance of the MAGIC telescopes during moonlight \citep{MAGIC_moon}. The systematic errors in the cut-off energy were estimated from the values of $E_c$ obtained when modifying the absolute light scale by $\pm 15\%$.

For the $Fermi$-LAT analysis, a broken power-law function with normalisation N$_{\rm o}=(8.0\pm0.4)\times10^{-12}$~ $\text{TeV}^{-1} \text{cm}^{-2} \text{s}^{-1}$, indices ${\Gamma_1=0.90\pm0.08}$ and ${\Gamma_2=2.37\pm0.04}$ is obtained and showed in Fig. \ref{fig:SED}, blue solid squares. The light gray shaded area shows the statistical errors of the obtained broken power-law fit whereas the dark one marks the uncertainty coming from the imperfectness in the Galactic diffuse emission modelling, dominating the Cas A flux uncertainties at low energies. The later were obtained by modifying the galactic diffuse flux by $\pm$6\%. Note that the systematic error due to the diffuse background is greatly reduced above 300 MeV. 

\section{Discussion}

MAGIC observations of the youngest GeV- and TeV-bright known SNR have allowed us to obtain the most precise spectrum of Cas A to date,  extending previous results obtained with Cherenkov instruments up to $\sim$10 TeV. In the MAGIC energy range, the spectrum is best-fit with a power-law with exponential cut-off function with index $\sim$2.4 and an energy cut-off at $E_c\sim$3.5~TeV. These findings provide a crucial insight into the acceleration processes in one of the most prominent non-thermal objects in our Galaxy. 

We also analysed more than 8 years of LAT data and obtained a spectrum that confirms the one by \cite{yuan_2013}. Below $\sim$1~GeV Cas A shows a hard spectrum with index $\sim$0.9. 
Above a few GeV, the spectrum measured with $Fermi$-LAT falls quickly with a photon index of $\sim$2.37, which is compatible within errors with the one measured using the MAGIC telescopes. 
%


To investigate the underlying population of particles, we have used the radiative code and Markov Chain Monte Carlo fitting routines of $naima$\footnote{https://github.com/zblz/naima} \citep{Zabalza_2015}, deriving the present-age particle distribution. The code uses the parametrisation of neutral pion decay by \cite{Kafexhiu_2014}, the parametrization of synchrotron radiation by \cite{NaimaSyn} and the analytical approximations to IC up-scattering of blackbody radiation and non-thermal bremsstrahlung developed by \cite{Khangulyan_2014} and \cite{Baring_1999}, respectively.

\begin{figure*}
\includegraphics[width=2.\columnwidth]{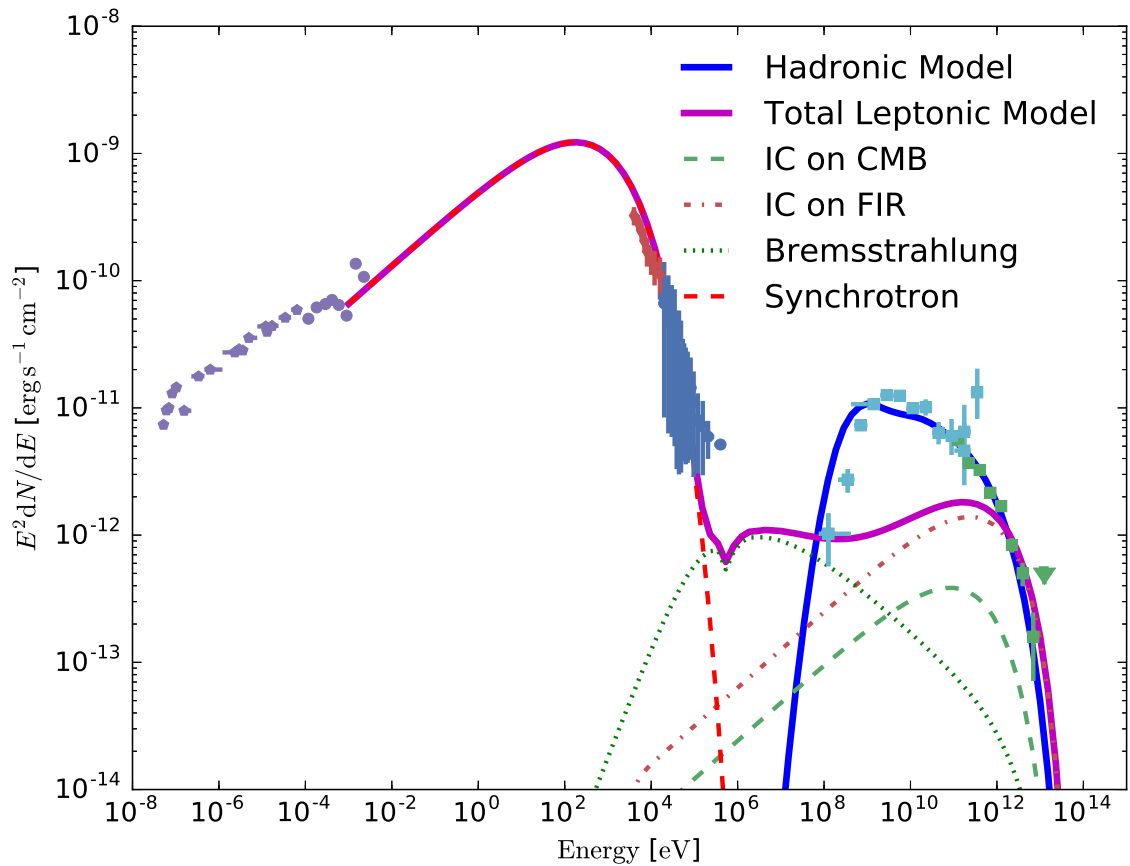}
\caption{Multi-wavelength SED of Cas~A. The different lines show the result of fitting the measured energy fluxes using $naima$ and assuming a leptonic or a hadronic origin of the GeV and TeV emission.}\label{fig:SEDModel}
\end{figure*}

We first considered the possibility that the gamma-ray emission was originated by an electron population, described with a power-law with an exponential cut-off function, producing Bremsstrahlung and IC radiation in the gamma-ray range, and synchrotron radiation at lower energies.
The photon fields that contribute to the inverse Compton component are the ubiquitous 2.7 K cosmic microwave background (CMB) and the large far infrared (FIR) field measured in Cas A, with a value of $\sim$2 eV/cm3 at 100 keV. Fixing the photon field to this value, we can obtain the highest possible density of electrons allowed by the VHE flux. Then we can constrain the maximum magnetic field for which the synchrotron radiation produced by the derived population does not exceed the radio and X-ray measurements\footnote{This constraint is due to the fact that, as reported in section~\ref{sect:intro}, several emission regions, likely associated to different particle populations, were identified at those wavelengths.}. The multi-wavelength SED is shown in Fig. \ref{fig:SEDModel}, with the radio emission displayed in purple dots \citep{Lastochkin_1963,Medd_1965,Allen_1967,Parker_1968,Braude_1969,Hales_1995,Planck_2014}, soft SUZAKU X-rays are marked in red \citep{suzaku_2009} and hard INTEGRAL X-rays in blue \citep{integral_2016}. In the gamma-ray regime, the LAT points are shown in cyan and the MAGIC ones in green. The MAGIC points can be described by an electron population with amplitude at 1~TeV of 2$\cdot$10$^{34}$eV$^{-1}$, spectral index 2.4 and cut-off energy at 8~TeV up-scattering the FIR (brown dash-dot line) and the CMB photons (green dashed line). The comparison with the low energy part of the SED constraints the magnetic field to B$\lessapprox$180~$\mu$G. The resulting emission from the leptonic model is shown in Fig. \ref{fig:SEDModel}. 
A relatively low magnetic field and a large photon field could be fulfilled in a reverse shock evolving in a thin and clumpy ejecta medium which provides a moderate amplification of the magnetic field and large photon fields in the clumps which are observed as optical knots.
The same population of electrons would unavoidably produce Bremsstrahlung radiation below a few GeV  (see green dotted line in Fig. \ref{fig:SEDModel}\footnote{Note that the structure in the spectral shape around 2~MeV is due to the transition between the two asymptotic regimes described in \cite{Baring_1999}, used in the \textit{naima} code.}). The emission observed with $Fermi$ LAT at the lowest energies constrain the density to n$\sim$1~cm$^{-3}$, still compatible with the smooth ejecta density \citep{Micelotta_2016}. 
The model is generally compatible with the X-ray points and with MAGIC spectrum above a few TeV, it is consistent with the radio measurements, but fails to reproduce the $\gamma$-ray spectrum between 1 GeV and 1 TeV, being a factor 2-3 below the measured LAT spectrum. In addition, to accommodate a magnetic field of the order of $\sim$1~mG, as reported in \cite{Uchiyama_2008}, the amplitude of the electron spectrum would need to be decreased at least by a factor 100, rendering a negligible IC contribution at the highest energies. 

Indeed the GeV-TeV emission of Cas A is usually attributed to accelerated protons. Assuming a population of CRs characterised with a power-law function with an exponential cut-off to fit the gamma-ray data from 60 MeV to 15 TeV, and a target density of 10 cm$^{-3}$ \citep{Laming_2003}. The proton spectrum is best-fit with a hard index of 2.21 and an exponential cut-off energy of 12~TeV, which implies a modest acceleration of CRs to VHE, well below the energy needed to explain the CR $knee$. The proton energy above 1 TeV is 5.1$\cdot$10$^{48}$ erg, which is only $\sim$0.2\% of the estimated explosion kinetic energy of ${\rm E_{\rm sn}} = 2\cdot10^{51}$ erg \citep{Laming_2003}. The total energy stored in protons above 100~MeV amounts to $9.9\cdot10^{49}$ erg.

The flat spectral index is in agreement with the standard theory of diffuse shock acceleration, but the low cut-off energy implies that Cas A is an extremely inefficient in the acceleration of CRs at the present moment. The characteristic maximum energy of these accelerated protons can be expressed, for standard parallel shock acceleration efficiency (see e. g. \citealt{Lagage_1983}), as: 

\begin{equation}
E^p_c \simeq 450 (\frac{B}{1~\rm {mG}})(\frac{t_0}{100~{\rm yr}})(\frac{u_s}{3000~{\rm km/s}})^2\eta^{-1}~ {\rm TeV},
\end{equation}

where $u_s\sim10^3~{\rm km/s}$ is the speed of the forward shock, $t_0\sim330~{\rm yrs}$ is the age of Cas A and $\eta\ge1$ is the acceleration efficiency (the ratio of the mean free path of a particle to its gyroradius), which is $\sim$1 in the Bohm diffuse regime. Even assuming a magnetic field as low as a few tens of $\mu$G, a poor acceleration efficiency $\eta\gg$10 has to be invoked to accommodate the low cut-off energy found. Alternatively, Cas A may also be located in a very diffusive region of the Galaxy, resulting in a very fast escape of protons of TeV and higher energies.

\section{Conclusions}

We report for the first time in VHE, observational evidence for the presence of a cut-off in the VHE spectrum of Cas A. The spectrum measured with the MAGIC telescopes can be described with an EPWL with a cut-off at $\sim$3.5 TeV, which is preferred over a PWL scenario with 4.6$\sigma$ significance. This result implies that even if all the TeV emission was of hadronic origin, Cas A could not be a PeVatron at its present age.

Several emission regions must be active to explain the radio, X-ray, GeV and TeV emission of Cas A. A purely leptonic model cannot explain the GeV-TeV spectral shape derived using LAT and MAGIC data, as previously suggested based on observations at lower energies 
\citep{Atoyan_2000_1, Atoyan_2000_2,Zirakashvili_2014,Saha_2014}. A leptonic population is undoubtedly necessary to explain the emission at radio and X-ray energies. Indeed the bright steep-spectrum radio knots and the bright radio ring, demand an average magnetic field of $\sim$1 mG \citep{Vink_2003}, whereas the faint plateau surrounding Cas A, seen in Chandra continuum images, is consistent with a lower magnetic field, which might contribute to the observed emission above 1~TeV. 

However, the bulk of the HE and VHE $\gamma$-rays must be of hadronic origin. 
Cas A is most likely accelerating CRs, although to a rather low energy of a few TeV. Even if some leptonic contribution at VHE produced by IC cannot be excluded, this would not affect our conclusion that acceleration in Cas A falls short of the energies of the $knee$ of the CR spectrum.

A detailed study of the cut-off shape is crucial to understand the reason behind this low acceleration efficiency, displaying different characteristics if due to escape of CRs, to the maximum energy of the accelerated CRs, or some other mechanism. Observations with the future Cherenkov Telescope Array (CTA, \citealt{Actis_2011}), with a superior angular resolution and sensitivity, will allow detailed spectroscopic investigation on the cut-off regime, providing new insights on the acceleration processes in Cas A.

\section*{Acknowledgements}
%
%

We would like to thank the Instituto de Astrof\'{\i}sica de Canarias for the excellent working conditions at the Observatorio del Roque de los Muchachos in La Palma. The financial support of the German BMBF and MPG, the Italian INFN and INAF, the Swiss National Fund SNF, the ERDF under the Spanish MINECO (FPA2015-69818-P, FPA2012-36668, FPA2015-68378-P, FPA2015-69210-C6-2-R, FPA2015-69210-C6-4-R, FPA2015-69210-C6-6-R, AYA2015-71042-P, AYA2016-76012-C3-1-P, ESP2015-71662-C2-2-P, CSD2009-00064), and the Japanese JSPS and MEXT is gratefully acknowledged. This work was also supported by the Spanish Centro de Excelencia ``Severo Ochoa'' SEV-2012-0234 and SEV-2015-0548, and Unidad de Excelencia ``Mar\'{\i}a de Maeztu'' MDM-2014-0369, by the Croatian Science Foundation (HrZZ) Project 09/176 and the University of Rijeka Project 13.12.1.3.02, by the DFG Collaborative Research Centers SFB823/C4 and SFB876/C3, and by the Polish MNiSzW grant 745/N-HESS-MAGIC/2010/0.


\bibliographystyle{mnras}

\begin{thebibliography}{3}
\expandafter\ifx\csname natexlab\endcsname\relax\def\natexlab#1{#1}\fi

\bibitem[{Abdo {et~al.}(2010)Abdo {et~al.} (the $Fermi$-LAT coll.)}]{abdo_2010}
Abdo, A. A., Ackermann, M., Ajello, M., {et~al.}, 2010, ApJL, 710, L92.



\bibitem[Ackermann et al.(2013)]{ackermann_2013} Ackermann, M., Ajello, M., Allafort, A., et al.\ 2013, Science, 339, 807 

\bibitem[Actis et al.(2011)]{Actis_2011} Actis, M., Agnetta, G., Aharonian, F., et al.\ 2011, Experimental Astronomy, 32, 193 

\bibitem[{Aharonian {et~al.}(2001)Aharonian {et~al.} (the HEGRA coll.)}]{aharonian_2001}
Aharonian, F., Akhperjanian, A., Barrio, J., {et~al.}, 2001, A\&A, 370, 112.

\bibitem[Aharonian et al.(2010)]{NaimaSyn} Aharonian, F.~A., Kelner, S.~R., \& Prosekin, A.~Y.\ 2010, \prd, 82, 043002

\bibitem[{Aharonian(2013)}]{aharonian_2013}
Aharonian, F., 2013, Astrop. Phys., 43, 71.

\bibitem[{Albert {et~al.}(2007)Albert {et~al.} (the MAGIC coll.)}]{albert_2007}
Albert, J., Aliu, E., Anderhub, H., {et~al.}, 2007, A\& A, 474, 937.

\bibitem[{{Aleksi{\'c}} {et~al.}(2016){Aleksi{\'c}}, {Ansoldi}, {Antonelli},
  {Antoranz}, {Babic}, {Bangale}, {Barcel{\'o}}, {Barrio}, {Becerra
  Gonz{\'a}lez}, {Bednarek}, {Bernardini}, {Biasuzzi}, {Biland}, {Bitossi},
  {Blanch}, {Bonnefoy}, {Bonnoli}, {Borracci}, {Bretz}, {Carmona}, {Carosi},
  {Cecchi}, {Colin}, {Colombo}, {Contreras}, {Corti}, {Cortina}, {Covino}, {Da
  Vela}, {Dazzi}, {De Angelis}, {De Caneva}, {De Lotto}, {de O{\~n}a Wilhelmi},
  {Delgado Mendez}, {Dettlaff}, {Dominis Prester}, {Dorner}, {Doro}, {Einecke},
  {Eisenacher}, {Elsaesser}, {Fidalgo}, {Fink}, {Fonseca}, {Font}, {Frantzen},
  {Fruck}, {Galindo}, {Garc{\'{\i}}a L{\'o}pez}, {Garczarczyk}, {Garrido
  Terrats}, {Gaug}, {Giavitto}, {Godinovi{\'c}}, {Gonz{\'a}lez Mu{\~n}oz},
  {Gozzini}, {Haberer}, {Hadasch}, {Hanabata}, {Hayashida}, {Herrera},
  {Hildebrand}, {Hose}, {Hrupec}, {Idec}, {Illa}, {Kadenius}, {Kellermann},
  {Knoetig}, {Kodani}, {Konno}, {Krause}, {Kubo}, {Kushida}, {La Barbera},
  {Lelas}, {Lemus}, {Lewandowska}, {Lindfors}, {Lombardi}, {Longo},
  {L{\'o}pez}, {L{\'o}pez-Coto}, {L{\'o}pez-Oramas}, {Lorca}, {Lorenz},
  {Lozano}, {Makariev}, {Mallot}, {Maneva}, {Mankuzhiyil}, {Mannheim},
  {Maraschi}, {Marcote}, {Mariotti}, {Mart{\'{\i}}nez}, {Mazin}, {Menzel},
  {Miranda}, {Mirzoyan}, {Moralejo}, {Munar-Adrover}, {Nakajima}, {Negrello},
  {Neustroev}, {Niedzwiecki}, {Nilsson}, {Nishijima}, {Noda}, {Orito},
  {Overkemping}, {Paiano}, {Palatiello}, {Paneque}, {Paoletti}, {Paredes},
  {Paredes-Fortuny}, {Persic}, {Poutanen}, {Prada Moroni}, {Prandini},
  {Puljak}, {Reinthal}, {Rhode}, {Rib{\'o}}, {Rico}, {Rodriguez Garcia},
  {R{\"u}gamer}, {Saito}, {Saito}, {Satalecka}, {Scalzotto}, {Scapin},
  {Schultz}, {Schlammer}, {Schmidl}, {Schweizer}, {Shore}, {Sillanp{\"a}{\"a}},
  {Sitarek}, {Snidaric}, {Sobczynska}, {Spanier}, {Stamerra}, {Steinbring},
  {Storz}, {Strzys}, {Takalo}, {Takami}, {Tavecchio}, {Tejedor}, {Temnikov},
  {Terzi{\'c}}, {Tescaro}, {Teshima}, {Thaele}, {Tibolla}, {Torres}, {Toyama},
  {Treves}, {Vogler}, {Wetteskind}, {Will}, \& {Zanin}}]{Magic_performanceII}
{Aleksi{\'c}}, J., {Ansoldi}, S., {Antonelli}, L.~A., {et~al.} 2016,
  Astroparticle Physics, 72, 76
  
\bibitem[Allen \& Barrett(1967)]{Allen_1967} Allen, R.~J., \& Barrett, A.~H.\ 1967, \apj, 149, 1

\bibitem[Anderson et al.(1991)]{Anderson_1991} Anderson, M., Rudnick, L., Leppik, P., Perley, R., \& Braun, R.\ 1991, \apj, 373, 146 

\bibitem[Anderson \& Rudnick(1995)]{Anderson_1995} Anderson, M.~C., \& Rudnick, L.\ 1995, \apj, 441, 307 

\bibitem[Atoyan et al.(2000a)]{Atoyan_2000_1} Atoyan, A.~M., Aharonian, F.~A., Tuffs, R.~J., \& V{\"o}lk, H.~J.\ 2000a, \aap, 355, 211 

\bibitem[Atoyan et al.(2000b)]{Atoyan_2000_2} Atoyan, A.~M., Tuffs, R.~J., Aharonian, F.~A., \& V{\"o}lk, H.~J.\ 2000b, \aap, 354, 915 

\bibitem[Baring et al.(1999)]{Baring_1999} Baring, M.~G., Ellison, D.~C., Reynolds, S.~P., Grenier, I.~A., \& Goret, P.\ 1999, \apj, 513, 311 

\bibitem[{Bell(2004)Bell}]{bell_2004}
Bell, A.R., 2004, MNRAS, 353, 550.

\bibitem[Bell et al.(2013)]{bell_2013_2} Bell, A.~R., Schure, K.~M., Reville, B., \& Giacinti, G.\ 2013, \mnras, 431, 415 

\bibitem[{Bell(2013)Bell}]{bell_2013}
Bell, A.R., 2013, Astrop. Phys., 43, 56.

\bibitem[Berezhko et al.(2003)]{Berezhko_2003} Berezhko, E.~G., P{\"u}hlhofer, G., \& V{\"o}lk, H.~J.\ 2003, \aap, 400, 971 

\bibitem[Braude et al.(1969)]{Braude_1969} Braude, S.~Y., Lebedeva, O.~M., Megn, A.~V., Ryabov, B.~P., \& Zhouck, I.~N.\ 1969, \mnras, 143, 289


\bibitem[DeLaney \& Rudnick(2003)]{DeLaney_2003} DeLaney, T., \& Rudnick, L.\ 2003, \apj, 589, 818 

\bibitem[Fomin et al.(1994)]{Fomin} Fomin, V.~P., Stepanian, A.~A., Lamb, R.~C., et al.\ 1994, Astroparticle Physics, 2, 137


\bibitem[{Gaisser(1991)Gaisser}]{gaisser_1991}
Gaisser, T.K., Cosmic Rays and Particle Physics, Cambridge University Press,
1991.


\bibitem[{Ginzburg(1964)Ginzburg and Syrovatskii}]{ginzburg_1964}
Ginzburg, V.L. and Syrovatskii, S.I., The Origin of Cosmic Rays, Macmillan, New York,
1964.

\bibitem[Gotthelf et al.(2001)]{Gotthelf_2001} Gotthelf, E.~V., Koralesky, B., Rudnick, L., et al.\ 2001, \apjl, 552, L39 

\bibitem[Grefenstette et al.(2015)]{Grefenstette_2015} Grefenstette, B.~W., Reynolds, S.~P., Harrison, F.~A., et al.\ 2015, \apj, 802, 15 


\bibitem[Hales et al.(1995)]{Hales_1995} Hales, S.~E.~G., Waldram, E.~M., Rees, N., \& Warner, P.~J.\ 1995, \mnras, 274, 447

\bibitem[Helder \& Vink(2008)]{Helder_2008} Helder, E.~A., \& Vink, J.\ 2008, \apj, 686, 1094-1102 

\bibitem[{Holder {et~al.}(2016)Holder for the VERITAS coll.}]{holder_2016}
Holder, J. for the VERITAS coll., 2016, in Proc. of the 6th International Symposium on High-Energy Gamma-Ray 
Astronomy (Gamma2016), Heidelberg, Germany, and arXiv:1609.02881 [astro-ph].

\bibitem[Kafexhiu et al.(2014)]{Kafexhiu_2014} Kafexhiu, E., Aharonian, F., Taylor, A.~M., \& Vila, G.~S.\ 2014, \prd, 90, 123014 


\bibitem[Khangulyan et~al.(2014)]{Khangulyan_2014}Khangulyan, D., Aharonian, F. A., \& Kelner, S. R. 2014, ApJ, 783, 100

\bibitem[Lagage \& Cesarsky(1983)]{Lagage_1983} Lagage, P.~O., \& Cesarsky, C.~J.\ 1983, \aap, 125, 249 

\bibitem[Laming \& Hwang(2003)]{Laming_2003} Laming, J.~M. \& Hwang, U. 2003, ApJ, 597, 347

\bibitem[Lastochkin et al.(1963)]{Lastochkin_1963} Lastochkin, V. P., Porfriev, V. A., Stankevich, K. S., Troitsky, V. S., Kholodilov, N. N., \& Tseitlin, N. M.\ 1963, Radiophysica (U.S.S.R.), 6, 629

\bibitem[Maeda et al.(2009)]{suzaku_2009} Maeda, Y., Uchiyama, Y., Bamba, A., et al.\ 2009, \pasj, 61, 1217 

\bibitem[MAGIC Collaboration (2017)]{MAGIC_moon} MAGIC Collaboration: Ahnen, M.~L., Ansoldi, S., et al.\ 2017, arXiv:1704.00906

\bibitem[Medd \& Ramana(1965)]{Medd_1965} Medd, W.~J., \& Ramana, K.~V.~V.\ 1965, \apj, 142, 383

\bibitem[Micelotta et al.(2016)]{Micelotta_2016} Micelotta, E.~R., Dwek, E., \& Slavin, J.~D.\ 2016, \aap, 590, A65

\bibitem[Mezger et al.(1986)]{Mezger_1986} Mezger, P.~G., Tuffs, R.~J., Chini, R., Kreysa, E., \& Gemuend, H.-P.\ 1986, \aap, 167, 145 

\bibitem[Morse et al.(2004)]{Morse_2004} Morse, J.~A., Fesen, R.~A., Chevalier, R.~A., et al.\ 2004, \apj, 614, 727 

\bibitem[Parker(1968)]{Parker_1968} Parker, E.~A.\ 1968, \mnras, 138, 407

\bibitem[Patnaude \& Fesen(2007)]{Patnaude_2007} Patnaude, D.~J., \& Fesen, R.~A.\ 2007, \aj, 133, 147 

\bibitem[O'C.~Drury(2014)]{Drury_2014} O'C.~Drury, L.\ 2014, arXiv:1412.1376

\bibitem[Planck Collaboration(2014)]{Planck_2014} Planck Collaboration, Ade, P.~A.~R., Aghanim, N., et al.\ 2014, \aap, 571, A28

\bibitem[{Reed {et~al.}(1995)Reed, Hester, Fabian, Winkler}]{reed_1995}
Reed, J. E., Hester, J. J., Fabian, A. C., \& Winkler, P. F. 1995, ApJ,
440, 706.


\bibitem[Rolke et al.(2005)]{Rolke_2005} Rolke, W.~A., L{\'o}pez, A.~M., \& Conrad, J.\ 2005, Nuclear Instruments and Methods in Physics Research A, 551, 493 

\bibitem[Saha et al.(2014)]{Saha_2014} Saha, L., Ergin, T., Majumdar, P., Bozkurt, M., \& Ercan, E.~N.\ 2014, \aap, 563, A88 

\bibitem[Siegert et al.(2015)]{Siegert_2015} Siegert, T., Diehl, R., Krause, M.~G.~H., \& Greiner, J.\ 2015, \aap, 579, A124 

\bibitem[Uchiyama \& Aharonian(2008)]{Uchiyama_2008} Uchiyama, Y., \& Aharonian, F.~A.\ 2008, \apjl, 677, L105 

\bibitem[Vink \& Laming(2003)]{Vink_2003} Vink, J., \& Laming, J.~M.\ 2003, \apj, 584, 758 

\bibitem[Wang \& Li(2016)]{integral_2016} Wang, W., \& Li, Z.\ 2016, \apj, 825, 102 

\bibitem[{Yuan {et~al.}(2013)Yuan, Funk, Johannesson, Lande, Tibaldo and Uchiyama}]{yuan_2013}
Yuan Y., Funk S., Johannesson G., Lande J., Tibaldo L. \& Uchiyama Y., 2013, ApJ 779, 117.

\bibitem[Zabalza(2015)]{Zabalza_2015} Zabalza, V. 2015, in Proceedings of the 34th International Cosmic Ray
Conference (ICRC 2015), in press, arXiv:1509.03319

\bibitem[{Zanin {et~al.}(2013)Zanin, Carmona, Sitarek, Colin, \&
  Frantzen}]{zanin2013mars}
Zanin, R., Carmona, E., Sitarek, J., Colin, P., \& Frantzen, K. 2013, in Proc.
  of the 33st International Cosmic Ray Conference, Rio de Janeiro, Brasil

\bibitem[Zirakashvili et al.(2014)]{Zirakashvili_2014} Zirakashvili, V.~N., Aharonian, F.~A., Yang, R., O{\~n}a-Wilhelmi, E., \& Tuffs, R.~J.\ 2014, \apj, 785, 130 


\end{thebibliography}

\vspace*{0.5cm}

\noindent
$^{1}$ {ETH Zurich, CH-8093 Zurich, Switzerland} \\
$^{2}$ {Universit\`a di Udine, and INFN Trieste, I-33100 Udine, Italy} \\
$^{3}$ {INAF - National Institute for Astrophysics, viale del Parco Mellini, 84, I-00136 Rome, Italy} \\
$^{4}$ {Universit\`a di Padova and INFN, I-35131 Padova, Italy} \\
$^{5}$ {Croatian MAGIC Consortium, Rudjer Boskovic Institute, University of Rijeka, University of Split - FESB, University of Zagreb - FER, University of Osijek,Croatia} \\
$^{6}$ {Saha Institute of Nuclear Physics, 1/AF Bidhannagar, Salt Lake, Sector-1, Kolkata 700064, India} \\
$^{7}$ {Max-Planck-Institut f\"ur Physik, D-80805 M\"unchen, Germany} \\
$^{8}$ {Universidad Complutense, E-28040 Madrid, Spain} \\
$^{9}$ {Inst. de Astrof\'isica de Canarias, E-38200 and Universidad de La Laguna, Dpto. Astrof\'isica, E-38206 La Laguna, Tenerife, Spain} \\
$^{10}$ {University of \L\'od\'z, PL-90236 Lodz, Poland} \\
$^{11}$ {Deutsches Elektronen-Synchrotron (DESY), D-15738 Zeuthen, Germany } \\
$^{12}$ {Humboldt University of Berlin, Institut f\"ur Physik Newtonstr. 15, 12489 Berlin Germany},\\
$^{13}$ {Institut de Fisica d'Altes Energies (IFAE), The Barcelona Institute of Science and Technology, Campus UAB, 08193 Bellaterra (Barcelona), Spain} \\
$^{14}$ {Universit\`a  di Siena, and INFN Pisa, I-53100 Siena, Italy} \\
$^{15}$ {Institute for Space Sciences (CSIC/IEEC), E-08193 Barcelona, Spain} \\
$^{16}$ {Technische Universit\"at Dortmund, D-44221 Dortmund, Germany} \\
$^{17}$ {Universit\"at W\"urzburg, D-97074 W\"urzburg, Germany} \\
$^{18}$ {Finnish MAGIC Consortium: Tuorla Observatory, University of Turku and Astronomy Division, University of Oulu, Finnish Centre for Astronomy with ESO (FINCA), Finland} \\
$^{19}$ {Unitat de F\'isica de les Radiacions, Departament de F\'isica, and CERES-IEEC, Universitat Aut\`onoma de Barcelona, E-08193 Bellaterra, Spain} \\
$^{20}$ {Universitat de Barcelona, ICC, IEEC-UB, E-08028 Barcelona, Spain} \\
$^{21}$ {Japanese MAGIC Consortium, ICRR, The University of Tokyo, Department of Physics, Kyoto University, Tokai University, The University of Tokushima, Japan} \\
$^{22}$ {Inst. for Nucl. Research and Nucl. Energy, BG-1784 Sofia, Bulgaria} \\
$^{23}$ {Universit\`a di Pisa, and INFN Pisa, I-56126 Pisa, Italy} \\

\label{lastpage}

\end{document}